\documentclass[]{article}
\usepackage[utf8]{inputenc}
\usepackage{amssymb}
\usepackage{amsmath}
\usepackage{enumerate}
\newcommand{\D}{\,\mathrm{d}}
\newcommand{\pd}[2]{\ensuremath{\frac{\partial #1}{\partial #2}}}
\newcommand{\pdl}[2]{\ensuremath{\partial #1 / \partial #2}}
\newcommand{\pdc}[3]{\ensuremath{\left(\frac{\partial #1}{\partial #2}\right)_{#3}}}

\begin{document}
\title{Reciprocal Relations\\ in Dissipationless Hydrodynamics}
\author{Lev A.\,Melnikovsky\thanks{E-mail: leva@kapitza.ras.ru}\\
\\
\textit{P.L.\,Kapitza Institute for Physical Problems}\\
\textit{Russian Academy of Sciences, Moscow, Russia}}
\date{}
\maketitle

\begin{abstract}
Hidden symmetry in dissipationless terms of arbitrary hydrodynamics
equations is recognized. We demonstrate that all fluxes are generated by a
single function and derive conventional Euler equations using proposed
formalism.
\end{abstract}

I am enjoying the privilege to learn from Alexander Fedorovich Andreev
for about 20 years. One of the doors he opened for me gives
access to the power of hydrodynamics. Universal
derivation procedure for its equations (seemingly pioneered
by Landau) is based purely on the conservation laws and symmetry considerations.
In present paper I would like to paraphrase this procedure in a formal
way without specifying particular system.

Generic equations of hydrodynamics are contained \cite{LL6} in local conservation laws
\begin{equation}
\label{genhydro}
\dot{y}_\alpha+\pd{j^k_\alpha}{x^k}=0.
\end{equation}
Here and below Latin superscripts are used for 3D space coordinates and 
Greek subscripts enumerate the integrals of motion; summation
over repeated indices is assumed.
Densities of these conserved quantities $y_\alpha$ form a complete set of
proper thermodynamic variables and unambiguously specify all local
equilibrium properties.

In dissipationless approximation an additional conservation law exists:
\begin{equation}
\label{entropy-conservation}
\dot{\sigma}+\pd{f^k}{x^k}=0,
\end{equation}
where $\sigma$ is the entropy density $\left(S=\int \sigma \D V\right)$.
The entropy density is a function of state 
\begin{equation}
\label{entropy-derivative}
\D \sigma=\pd{\sigma}{y_\alpha} \D y_\alpha.
\end{equation}

Equations \eqref{genhydro} have internal symmetry which is revealed if expressed
through the thermodynamically conjugate quantities (see \cite{LL5})
$Y_\alpha=-\pdl{\sigma}{y_\alpha}$.
The transformation from $y_\alpha$ to $Y_\alpha$ is invertible and one may use either
set of variables to characterize the state.

The fluxes $j^k_\alpha$ and $f^k$ generally depend on $Y_\alpha$ (or, equally, on $y_\alpha$)
and their spatial derivatives. In dissipationless approximation the dependence on spatial
derivatives is neglected. We may therefore substitute 
\eqref{entropy-derivative} in \eqref{entropy-conservation} and transform 
as follows
\begin{equation*}
0=\pd{\sigma}{y_\alpha} \dot{y}_\alpha +
\pd{f^k} {x^k}=
\left(Y_\alpha  \pd{j^k_\alpha}{Y_\beta} +
\pd{f^k}{Y_\beta}\right) \pd{Y_\beta} {x^k}.
\end{equation*}
This equation holds for arbitrary $\pdl{Y_\beta}{x^k}$ and the expression in parentheses
is identically zero:
\begin{equation*}
0=Y_\alpha  \pd{\mathbf{j}_\alpha}{Y_\beta} +
\pd{\mathbf{f}}{Y_\beta}=
\pd{}{Y_\beta}
\left(Y_\alpha  \mathbf{j}_\alpha
+
\mathbf{f}\right)-\mathbf{j}_\beta.
\end{equation*}
In other words, all fluxes $\mathbf{j}_\alpha$ are generated by a single function $\mathbf{g}$
\begin{equation}
\mathbf{j}_\alpha=
\pd{\mathbf{g}}{Y_\alpha}
\end{equation}
and the entropy flux $\mathbf{f}$ is its Legendre transform
\begin{equation}
\mathbf{f}=\mathbf{g}-Y_\alpha\mathbf{j}_\alpha.
\end{equation}

This completes the proof, that the matrix of derivatives $\pdl{j^k_\alpha}{Y_\beta}$ is symmetric
\begin{equation}
\pd{\mathbf{j}_\alpha}{Y_\beta}=
\pd{\mathbf{j}_\beta}{Y_\alpha}
\end{equation}
in agreement\footnote{Macroscopic reversibility
implies that $\mathbf{j}_\alpha$ and $Y_\alpha$ (or $y_\alpha$) behave oppositely
under time reversal.} with Onsager principle~\cite{onsager}.

The proof does not rely upon any specific property of the entropy itself.
In fact, equation \eqref{entropy-conservation} is just one
of local conservation laws and ``thermodynamically conjugate quantities''
$Y_\alpha$ could be defined with respect to any other integral of motion.
In practice it might be more convenient to use energy rather than entropy
to this end.

To illustrate this formalism, consider a classical ideal fluid. Energy per
unit volume $E$ is a function of other conserved quantities:
\begin{equation*}
\D E= T\D\sigma +\left(\mu-\frac{v^2}{2}\right)\D\rho + \mathbf{v}\D\mathbf{j},
\end{equation*}
where $\mu$ is chemical potential, $\rho$ is mass density,
$\mathbf{v}$ is fluid velocity,  and $\mathbf{j}$ is momentum density.
Conjugate variables are therefore $Y_\sigma=-T$, $Y_\rho=v^2/2-\mu$,
$Y_\mathbf{j}=-\mathbf{v}$. Due to the fluid isotropy, the generating function is
$\mathbf{g}=\mathbf{v}h(T,Y_\rho,v)$, where the scalar $h$ can be obtained
from identity $\mathbf{j}_\rho=\mathbf{j}=\rho\mathbf{v}$:
\begin{equation*}
\rho\mathbf{v}=
\pdc{\mathbf{g}}{Y_\rho}{T,\mathbf{v}}=
-\mathbf{v}\pdc{h}{\mu}{T,\mathbf{v}}.
\end{equation*}
Recalling the expression for the pressure differential $\D p = \sigma\D T +
\rho\D\mu$, we get $h=-p$. One can easily verify that the function
$\mathbf{g}=-p\mathbf{v}$ generates conventional Euler fluxes:
\begin{align*}
\mathbf{j}_\sigma=\pdc{\mathbf{g}}{Y_\sigma}{Y_\rho,Y_\mathbf{j}} &= \sigma \mathbf{v},\\
\Pi^{ik}=j^i_{j^k}=\pdc{g^i}{Y_{j^k}}{Y_\sigma,Y_\rho} &= p \delta^{ik}+\rho v^i v^k.
\end{align*}

\end{document}